\documentclass{nature}
\usepackage[utf8]{inputenc}
\usepackage[T1]{fontenc}
\usepackage{graphicx}
\usepackage{xcolor}
\usepackage{physics}
\usepackage{bm}
\usepackage{amsfonts}
\usepackage{soul}
\usepackage{hyperref}
\hypersetup{
    colorlinks=true, linkcolor=blue, filecolor=blue, urlcolor=blue,citecolor=blue,}

\usepackage{upgreek}

\makeatletter
\let\saved@includegraphics\includegraphics
\AtBeginDocument{\let\includegraphics\saved@includegraphics}
\renewenvironment*{figure}{\@float{figure}}{\end@float}
\makeatother
\title{Direct Observation and Optical Manipulation of Exciton-polariton Parametric Scattering Lasing in Temporal}
\author{%
Junxing Dong\textsuperscript{1,\#}, 
Si Shen\textsuperscript{2,\#,*}, 
Jingzhuo Wang\textsuperscript{1}, 
Lisheng Wang\textsuperscript{1}, 
Yifan Zhang\textsuperscript{1}, 
Huashan Li\textsuperscript{1,*}, 
Xianghu Wang\textsuperscript{2}, 
Wei Gao\textsuperscript{3}, 
Yongzheng Fang\textsuperscript{3} 
and Hai Zhu\textsuperscript{1,*}}
\DeclareUnicodeCharacter{2215}{/}
\begin{document}
\begin{spacing}{1.125}
\maketitle
\begin{scriptsize}
\begin{affiliations}
\item State Key Laboratory of Optoelectronic Materials and Technologies, School of Physics, Sun Yat-Sen University, Guangzhou 510275, China
\item School of Arts and Sciences, Shanghai Dianji University, Shanghai 200245, China
\item School of Materials Science and Engineering, Shanghai Institute of Technology, Shanghai, 201418, China 
\\
${\#}$ The authors contribute equally
\\
${*}$ Corresponding author 32188@sdju.edu.cn, lihsh25@mail.sysu.edu.cn, zhuhai5@mail.sysu.edu.cn
\end{affiliations}

\end{scriptsize}

\begin{abstract}
The hybrid light-matter character of exciton-polaritons gives rise to distinct polariton parametric scattering (PPS) process, which holds promise for frontier applications in polaritonic quantum devices. However, the stable excitation and coherent optical manipulation of PPS remain challenging due to scattering bottlenecks and rapid dephasing effect in polariton many-body systems. In this study, we first report the direct observation and optical amplification of non-degenerate intermode PPS lasing at room temperature (RT). The specific polariton branch of strong-coupled nanobelt planar microcavity is resonantly excited by a near-infrared (NIR) femtosecond laser via two-photon absorption (TPA) scheme, and the non-degenerate signal- and idler-states are stimulated. Angle-resolved dispersion patterns clearly reveal the evolution of the pump-, signal-, and idler-states under different excitation powers. Based on our self-constructed ultrafast femtosecond resonant optical trigger set-up, a selective enhancement and modulation of the signal-state is realized. Furthermore, the dynamic measurements of nonlinear signal-state enhancement process demonstrate a sub-picosecond response time (0.4ps), confirming its potential for ultrafast optical manipulation. Our work establishes a platform for exploring TPA-driven PPS laser and provides a novel optical modulation route for polariton-based optoelectronic devices.
\end{abstract}

\vspace{5mm}

Exciton-polaritons are quasi-particles generated under strong coupling between semiconductor excitons and microcavity photons. The unique low effective mass and strong nonlinearity of polaritons, making them a focus of intense research in fields including nonlinear optics and quantum information~\cite{Deng_2003, Deng_2010, Byrnes_2014}. Recently, plenty of quantum phenomena of polaritons involve Bose-Einstein condensation lasing~\cite{Kasprzak_2006, Christmann_2008, Kena_Cohen_2010, Lu_2012, Xu_2014, Su_2017, Chen_2020, Tang_2021, Freire_Fernandez_2024}, superfluidity~\cite{Amo_2009a, Kohnle_2011, Lagoudakis_2008}, and quantum vortices~\cite{Lagoudakis_2009, Tosi_2011, Barrat_2024, Yan_2025, Amo_2009b} have been demonstrated. Among various nonlinear processes, the microcavity PPS is of high interest. PPS is a typical three-wave mixing interaction process in polariton system, in which the signal- and idler-states are generated from pump-state under energy and momentum conservation~\cite{Ciuti_2000}. In the past decade, several mechanisms of PPS stimulated amplification have been reported~\cite{Savvidis_2000, Diederichs_2006, Xie_2012, Wu_2021, Tian_2022, Wang_2024}. On the basis of low-threshold, high coherence and optical tunability, PPS exhibits significant potential applications in all-optical logic gate and modulator for quantum compute and on-chip optical interconnects~\cite{Saba_2001, Kundermann_2003, Chen_2022a, Chen_2022b, Zhao_2022, Zhu_2025}.

Generally, non-radiative losses and rapid decoherence of excitons hinder the scattering efficiency of PPS~\cite{Savvidis_2000, Zhao_2022, Kuwata_Gonokami_1997}. Moreover, the strict energy and momentum conservation condition in PPS constrains the pump-state selection and excitation pathway~\cite{Deng_2003, Su_2017}. Therefore, the proper initiation and high conversion efficiency of PPS are crucial for ultrafast parametric control~\cite{Kundermann_2003, Chen_2022a, Chen_2022b}. To improve the intensity of signal-state, non-equilibrium resonant excitation scheme has been proposed~\cite{Wang_2024, Saba_2001, Kundermann_2003, Dasbach_2002}. However, the extreme complexity of pump-state configuration makes detection and control difficult~\cite{Wang_2024}. Up to date, the stable stimulation emission and coherent manipulation of the PPS process still remain challenging.

In this paper, we first report the direct probe and ultrafast manipulation of non-degenerate polariton PPS lasing at RT in a strong-coupled semiconductor nanobelt microcavity. The detailed evolution of PPS in lower polariton (LP) branches including the pump-, signal-, and idler-states are explored in angle-resolved dispersion patterns under resonant TPA excitation. The threshold of stimulated PPS lasing is about 1.1~GW/cm\textsuperscript{2}, and two-dimensional \textit{k}-space imaging further reveals the scattering mechanism underlying amplification process. In addition, an optical triggering scheme is designed to selectively enhance the signal-state with an amplification of 6.91~dB. The response time is extracted to be 0.4~ps in trigger process, confirming the ultrafast active modulation ability. Our results serve as a platform for direct probing of the signal- and idler-state in non-degenerate PPS, and offers a promising pathway for advancing polaritonic devices based on nonlinear PPS mechanism.

The schematic configuration of the planar microcavity and experimental setup presents a pair of symmetric DBRs with a ZnO nanobelt active gain medium, forming a vertical-cavity surface-emitting laser (VCSEL) structure (Fig.~1a). Scanning electron microscopy (SEM) image of the nanobelt reveals a well-defined rectangular cross-section and clear geometric boundaries (Fig.~1b). To excite the PPS, a mode-locked Ti:sapphire femtosecond laser with central wavelength of 810~nm (1.53~eV) is used as the pumping source. Photon energy of the laser is about half of the ground-state energy of LP\textsubscript{1} branch, thereby polaritons can be activated via resonant TPA effect. Meanwhile, the TPA excitation allows incident light beam to be detuned precisely from the high-reflectivity spectral region of DBRs, optimizing both excitation efficiency and signal collection~\cite{Wang_2024}. Based on our designed k-space modulation unit, the excitation laser can be focused onto the microcavity through a NIR objective, enabling selective excitation of the pump-state (LP\textsubscript{2}). On the basis of energy and momentum conservation, two polariton quasi-particles at pump-state generate signal-state at LP\textsubscript{1} branch and idler-state at LP\textsubscript{3} branch respectively. To achieve efficient detection of PPS, a near-ultraviolet (NUV) objective is adopted to confocally collect the transmitted spectra from the microcavity.

The photoluminescence (PL) spectrum of the bare nanobelt shows a prominent near band emission (NBE) peak near 3.3~eV (Fig.~1c)~\cite{Djurisic_2004, Schmidt_2010}. The reflectance properties of the DBRs exhibit a high-reflectivity band that covers the exciton emission peak, thereby providing spectral overlap conditions for strong exciton-photon coupling. Furthermore, the angle-resolved dispersion pattern of the microcavity PL clearly reveals the emission of LP\textsubscript{1}, LP\textsubscript{2}, LP\textsubscript{3} branch (Fig.~1d) that exhibits well-distributed energy bands and arranged polariton population in momentum space. These unique characteristics of the designed VCSEL provide a favorable physical route for the efficient coherent PPS process between different LP branches. The critical scattering positions that satisfy both energy and momentum conservation have been marked in the pattern.

The evolution of the intermode stimulated PPS process excited via TPA is measured at RT. At low excitation density (0.5~P\textsubscript{th}), the resonantly injected quasi-particles are primarily distributed at the pump-state (Fig.~2a) located on the LP2 branch at $k_{ \| } \sim 3.4\mu m ^ { - 1}$. Only a small fraction undergo Rayleigh scattering into signal-state (LP1) and idler-state (LP3)~\cite{Wu_2021, Baumberg_2000}, corresponding to adjacent momentum regions in k-space. As the particle density increases to threshold (Fig.~2b), the injection rate of signal-state polaritons exceeds the dissipation rate, leading to the accumulation of polaritons. Consequently, a significantly stimulated PPS lasing emerges on the LP1 branch along with population increase. Remarkably, rather than signal-state strictly forming at k\textsubscript{‖}~=~0, the PPS process occurs at a momentum position that best satisfies both energy and momentum conservation. The prominent signal-state emission around $k_{ \| } \sim 0.9\mu m ^ { - 1}$ is realized, with the corresponding idler-state located near $k_{ \| } \sim 5.9\mu m ^ { - 1}$. Above three states appear a robust momentum-conserving scattering triad ($2k_{\mathrm{pump}}=k_{\mathrm{signal}}+k_{\mathrm{idler}}$). At higher power (1.9~P\textsubscript{th}), a large number of polaritons accumulate in the stimulated scattering signal-state and the emission is dominated by the signal state (Fig.~2c). The corresponding PPS process is simulated using the generalized Gross-Pitaevskii (G-P) equation~\cite{Ciuti_2000, Wouters_2007} under excitation densities of 0.5, 1.0, and 1.9~P\textsubscript{th}, respectively (Fig.~2d-f). These theoretical simulations agree with the observed evolution of the PPS process, confirming the physical mechanism of the TPA-driven intermode PPS in our fabricated device (Supplementary Information).

To further reveal the intrinsic scheme and statistical characteristics of PPS, quantitative analysis of the emission intensity, linewidth, and energy shift in PPS is performed. The plot of the integrated signal-state intensity versus excitation power can be fitted with a nonlinear coefficient of 1.9 (Fig.~2g), which is consistent with the theoretical value of 2 expected for TPA process~\cite{Boyd_2008, Chen_2018}. As the power exceeds the threshold $P_{\mathrm{th}}$, the signal-state emission intensity shows a nonlinear increasing trend, indicating that the polariton system enters the stimulated PPS regime. The variations of pump- and idler-state emission intensities deviate from the double-logarithmic linear trend near $P_{\mathrm{th}}$, but the deviation is smaller than that of the signal-state, indicating that energy mainly accumulates in the signal-state (inset of Fig.~2g). The full width at half maximum (FWHM) of the three states are shown in Fig.~2h. With increasing excitation power, both the signal- and idler-state exhibit a clear narrowing of the linewidth, reflecting an enhancement in the system's coherence and particle correlation. The FWHM of the signal-state emission decreases to 8~meV at high excitation power, meanwhile the peak of signal-state displays an energy blueshift with increasing excited power (Fig.~2i). Notably, the initial energy separation between the three states is nearly equal, which satisfy the energy conservation relation $2E_{\mathrm{pump}} = E_{\mathrm{signal}} + E_{\mathrm{idler}}$. At high excitation power, both signal- and idler-state resulted from stimulated PPS exhibit the typical blueshift behavior. The slight energy change has been demonstrated to be related with enhanced particle correlation effect induced by Bose-Einstein statistics~\cite{Deng_2010, Byrnes_2014}. In contrast, the pump-state displays a redshift action due to the reconstruction of state density and the alteration of polariton quasi-particle occupation.

The two-dimensional (2D) \textit{k}-space images can explore the transition from Rayleigh scattering to nonlinear PPS in momentum space~\cite{Wu_2021}. As the excitation power increases from 0.4 to 2.0~P\textsubscript{th}, the dominant emission region gradually shifts from the pump-state to the signal-state (center point), clearly illustrating the evolution of microcavity PPS (Fig.~3a-d). The 2D spatial profiles of three states in \textit{k}-space exhibit notable divergent quasi-particles occupations. For the pump-state determined by the excitation pulse, it appears as a circular distribution. In contrast, the signal-state shows a smaller elliptical dot, while the idler-state tends to exhibit an arc-shaped emission. The distinctive features indicate that the signal- and idler-state are not directly excited by the resonant pump but originate from stimulated PPS. The coherent quasi-particle scattering occurs within a fan-shaped region predominantly, spanning an angular range of approximately 50\textdegree. It should be noted that the idler-state enhancement displays more prominently in the 2D \textit{k}-space pattern. This discrepancy arises because the angle-resolved setup collects spectra along a fixed \textit{k\textsubscript{y}} slice and thus does not capture the full integrated emission from the LP\textsubscript{3} branch. To clearly illustrate the geometric configuration of PPS process in \textit{k}-space, the three-dimensional representation of PPS emission at 2.0~P\textsubscript{th} is given (Fig.~3e), with the fan-shaped scattering region clearly annotated. It is evident that polaritons in pump-state bifurcate along two symmetric trajectories in \textit{k}-space, generating point-like signal-state in small-\textit{k} region and arc-shaped idler-state in high-\textit{k} region. The 2D \textit{k}-space measurements further validate the momentum evolution and phase-matching characteristics of stimulated PPS process.

Here, we further demonstrate the enhancement and modulation effects of the PPS process in strong-coupled device. The schematic of the optically triggered setup based on the resonant pump excitation is presented (Fig.~4a). A weaker pulse, serving as a resonant optical trigger, is introduced at a low angle ($k_{\parallel} \sim 0$) with the same photon energy as the pump pulse. The detailed optical path implementation is illustrated (Fig.~S1), and the power density of the trigger pulse is set as 0.3~GW/cm$^{2}$. Near the center of momentum space, the trigger stimulates the signal-state selectively without altering the excitation conditions of the pump pulse. The observed enhancement arises from seed polaritons injected by the trigger beam, which initiate coherent scattering processes along specific wavevector directions. Above route provide a feasible ultrafast control and amplification mechanism for the microcavity PPS emission.

The population of polaritons in k-space can be revealed directly through the comparison of the emission patterns with the optical triggered off and on (Fig.~4b-c). To highlight the distinction between the two conditions, the idler-state emission has been numerically amplified (×50), and the background contribution from trigger light has been subtracted. Under the trigger-on condition, a strong enhancement of signal-state intensity can be observed with the emission spot nearby \( k_{\parallel} = 0 \), indicating more efficient population of signal-state polaritons in small-momentum region via PPS. Meanwhile, the idler-state also exhibits a noticeable change of intensity with trigger, confirming the correlated response and coupling with the signal-state. The quantitative comparison of power-dependent emission intensities of signal- and idler-states with (without) the triggering pulse is present (Fig.~4d-e). The trigger pulse induces a significant amplification of 6.91~dB (2.81~dB) for signal-state (idler-state), while the thresholds of stimulated PPS are almost invariable. The enhancement behavior indicates that the optical trigger serves as a ``seed injection'', providing a localized perturbation that modifies the phase-matching conditions in momentum space. Consequently, the signal-state emission properties can be effectively manipulated, offering a powerful strategy for controlling PPS in strong-coupled microcavity.

In order to explore the mechanism of optical-trigger amplification, we analyzed the angular dependence and temporal dynamics of PPS with triggering (Fig.~5a). The enhanced signal-state emission exhibits a pronounced angular dependence of the trigger pulse. As the incidence angle of trigger pulse approaches $k_{\parallel} = 0$, the amplification of signal-state reaches the maximum. In the case of incidence angle shifts to the positive direction (toward the excitation pulse), the enhancement effect gradually attenuates. Conversely, when the incidence angle shifts toward negative $k_{\parallel}$ region, the signal-state emission intensity exhibits a reduction. Since the seed-injection inherently acts as a resonant amplification mechanism rather than suppression of scattering channels, the reduction can be attributed to polariton quenching and thermal dissipation effect that induced by the trigger beam.

For a coherent three-wave mixing process, the microcavity PPS is inherently governed by ultrafast dynamics. Here, the resonant triggering scheme is employed and serves as a critical probe to dissect the rapid interactions~\cite{Saba_2001, Zhao_2022, Zasedatelev_2019}. The temporal delay scan of signal-state emission with the trigger incident at $k_{\parallel} = 0.3~\mu\text{m}^{-1}$ and $-1.2~\mu\text{m}^{-1}$ are presented (Fig.~5b). These two angular positions are highlighted in Fig.~5a using distinct colors. The optimal enhancement angle locates at $k_{\parallel} = 0.3~\mu\text{m}^{-1}$ (purple curve), the temporal response exhibits a typical Gaussian waveform. The fitting gives a FWHM of 0.4~ps which is slightly broader than the excitation pulse (0.25~ps), indicating the presence of a sub-picosecond relaxation process in PPS. In contrast, no discernible response is observed within the scanning window (red curve) at the suppression point $k_{\parallel} = -1.2~\mu\text{m}^{-1}$. The observed intensity suppression can be attributed to the slower thermal relaxation process in solid materials. Since thermal cooling in many-body polariton systems typically occurs over nanosecond timescales~\cite{Tang_1993, Huang_2018}, no observable PPS emission decay is captured within the $\pm 250$~ps temporal range of our experiment setups. The normalized dispersion patterns of signal-state at two critical time points (Fig.~5c-d), corresponding to the markers on the purple curve in Fig~5b. In the mis-matching time regime (Fig.~5d), the trigger pulse and signal-state emission occupy two separate regions in k-space, resulting in two distinct emission lobes without spatial overlap. Hence, no nonlinear interaction between two discrete polariton states is formed through strong coupling scheme. At the optimal matching time point (Fig.~5e), the two signals spatially merge into a single emission region, leading to coherent spatiotemporal overlap. The spatial-temporal merging confirms that the resonant trigger induces wavefunction coherence between the injected seed polaritons and the signal-state. Consequently, directional injection of polaritons into the signal mode occurs within a sub-picosecond timescale, thereby enabling coherent enhancement of the stimulated PPS.

In conclusion, we have realized the non-degenerate PPS lasing at RT with the temporal resolved stimulated emission and optically triggered amplification. The innovative selected excitation of the pump-state is achieved via TPA effect, enabling a well-resolved stimulated PPS involving distinct signal- and idler-states. The resonant optical manipulation scheme is proposed, demonstrating enhanced PPS emission and verified phase-matching reconstruction, with a maximum amplification of 6.91 dB. In addition, dynamic measurements reveal that the amplification action of PPS exhibits an ultrafast temporal response (0.4 ps). The designed platform overcomes the detection bottleneck of high-momentum idler-states and phase-matching constraints for coherent PPS. Our results not only illustrate the intrinsic temporal action of nonlinear PPS but also promises the applications of ultrafast polariton optical source with coherent control and trigger of parametric oscillators.

\vspace{5mm}

\section*{Experimental Section}
\subsection{Fabrication of ZnO microcavity.}
Fabrication of ZnO microcavity: A tube furnace serves as the CVD platform, with zinc powder and O\textsubscript{2} gas acting as precursors for ZnO formation. Zn powder and a gold-coated quartz substrate are placed within the tube. Under mixed O\textsubscript{2}/Ar gas flow at 960\textdegree C, Zn vapor dissolved into molten Au droplets in the presence of the Ar/O\textsubscript{2} mixture, enabling the growth of two-dimensional ZnO nanobelt structures. The individual nanobelt exhibiting variable thicknesses is selectively identified through optical microscopy and dry-transferred onto bottom DBRs (10 pairs of Ta\textsubscript{2}O\textsubscript{5}/SiO\textsubscript{2} $\lambda/4$ film), deposited via electron-beam evaporation. Subsequently, the top DBRs is deposited onto the nanobelt, forming a Fabry–Pérot-type planar optical microcavity.
\subsection{Optical Characterization.}
For TPA excitation of PPS in this experiment, a mode-locked Ti:sapphire laser (Spectra-Physics Tsunami) is employed, with the femtosecond pulses of 810 nm. The repetition rate and pulse width of the pumping laser are 80 MHz and 250 fs, respectively. The excitation beam is focused onto the sample through an objective lens (Mitutoyo 50X, 0.42 NA NIR), while confocal signal collection is achieved using a counterpart objective lens (Mitutoyo 50X, 0.42 NA NUV). Post-collection signal is directed to a home-built angle-resolved micro-PL system for analysis, with angle-resolved spectra captured by a spectrometer equipped with a silicon charge-coupled device (Princeton Instruments SP2500, Acton). Moreover, a 325 nm He-Cd continuous-wave laser (Kimmon IK3301R-G) is used to excite the bare nanobelt. Temporal measurements incorporated a motorized translation platform (OptoSigma SHOT-702, SGSP20-85) serving as optical delay-line stage. More optical configuration details are provided in Supplementary Information.

\vspace{5mm}
\section*{Data Availability}
\noindent All data that supports the conclusions of this study are included in the article. The data presented in this study are available from the corresponding author upon reasonable request.

\vspace{5mm}
\section*{Acknowledgments} 
\noindent  We acknowledge financial support from the National Natural Science Foundation of China (No. U22A2073, 62474197, 52473244). Guangdong Basic and Applied Basic Research Foundation (2024A1515011536, 2025A1515011381). Guangzhou Basic and Applied Basic Research Foundation (2025A04J7142, 2025A04J4372). The Research Center for Magnetoelectric Physics of Guangdong Province (Grants 2024B0303390001), and the Guangdong Provincial Key Laboratory of Magnetoelectric Physics and Devices (No. 2022B1212010008).

\vspace{5mm}
\section*{Conflict of Interest} 
\noindent The authors declare no conflict of interest.

\vspace{5mm}

\section*{References}
\vspace{10mm}
\bibliographystyle{naturemag}
\bibliography{bib}

\end{spacing}

\clearpage{}
\newpage{}
\begin{figure}
     \centering
        \centering  \includegraphics[width=1\textwidth]{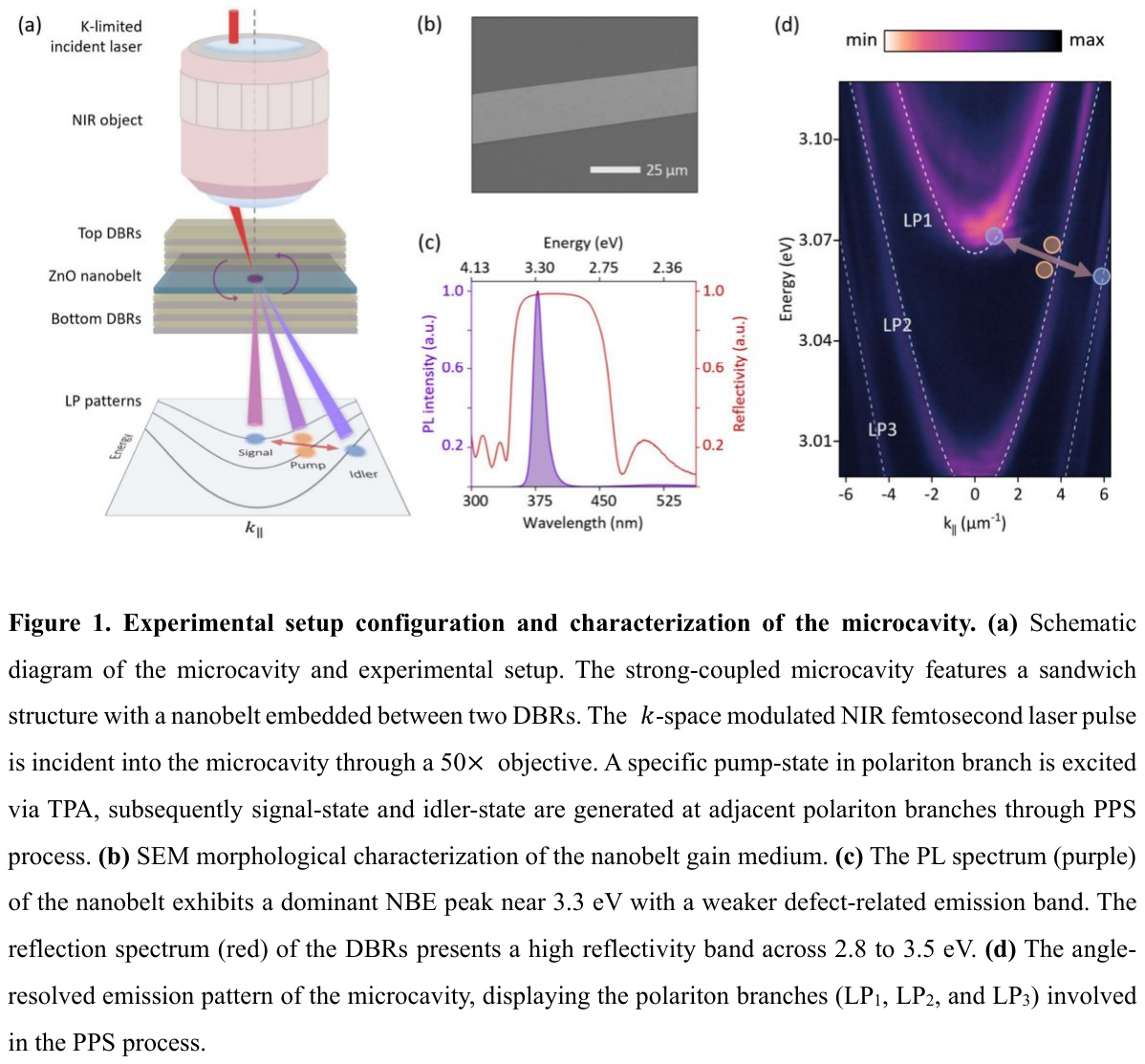}
    \label{fig_1}
\end{figure}

\clearpage{}
\newpage{}
\begin{figure}
     \centering
        \centering  \includegraphics[width=1\textwidth]{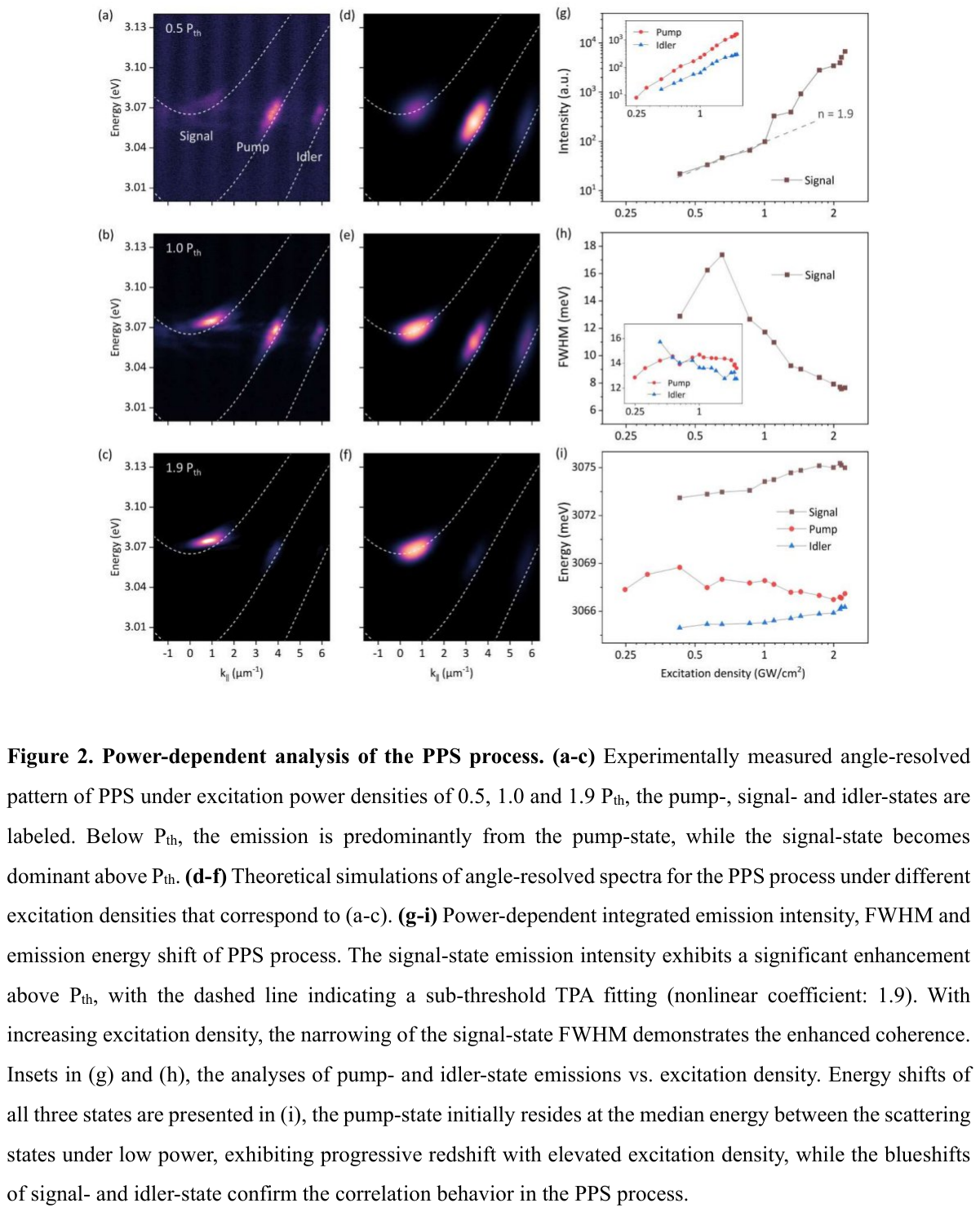}
    \label{fig_1}
\end{figure}

\clearpage{}
\newpage{}
\begin{figure}
     \centering
        \centering  \includegraphics[width=1\textwidth]{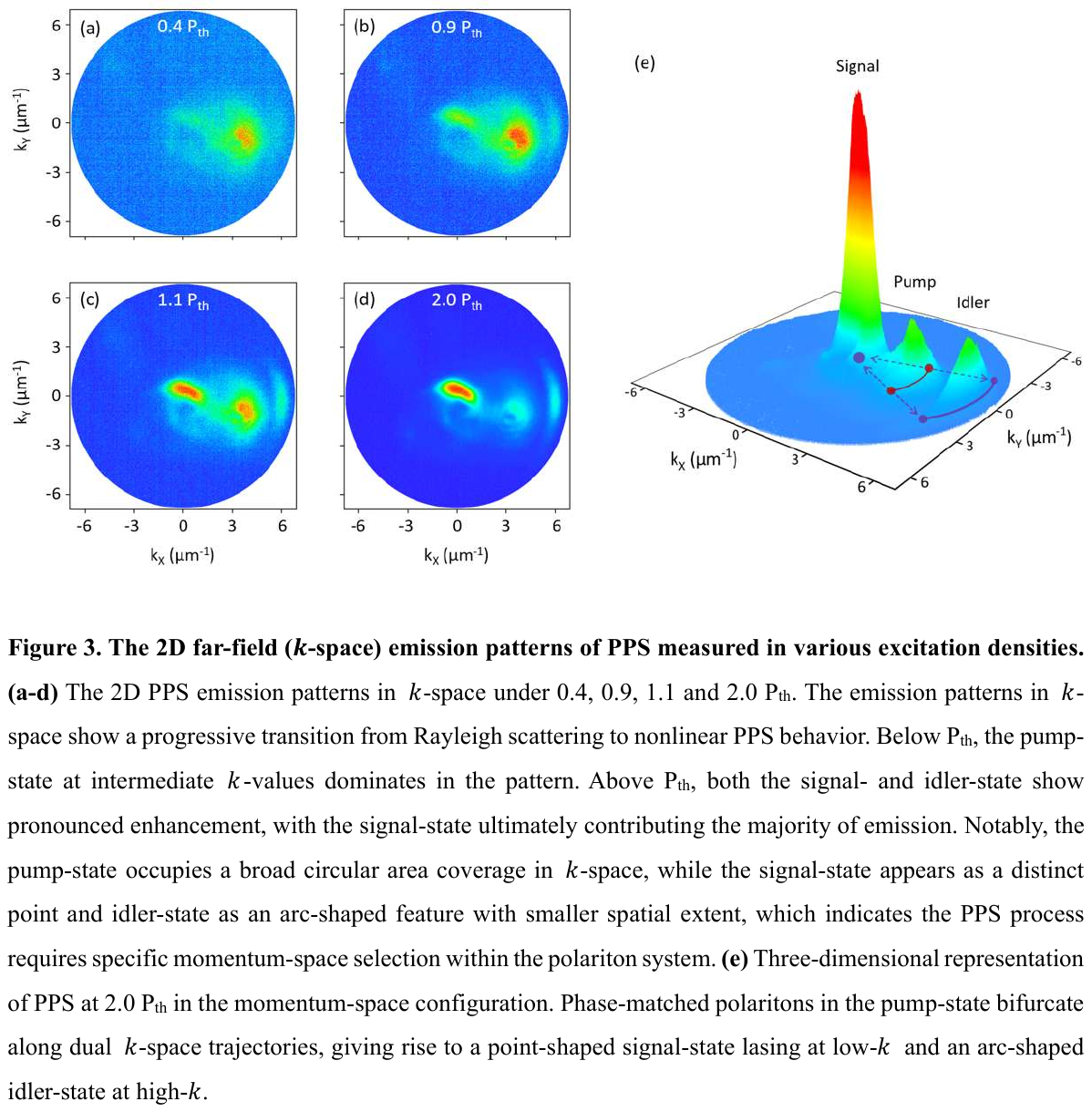}
    \label{fig_2}
\end{figure}

\clearpage{}
\newpage{}
\begin{figure}
     \centering
        \centering  \includegraphics[width=1\textwidth]{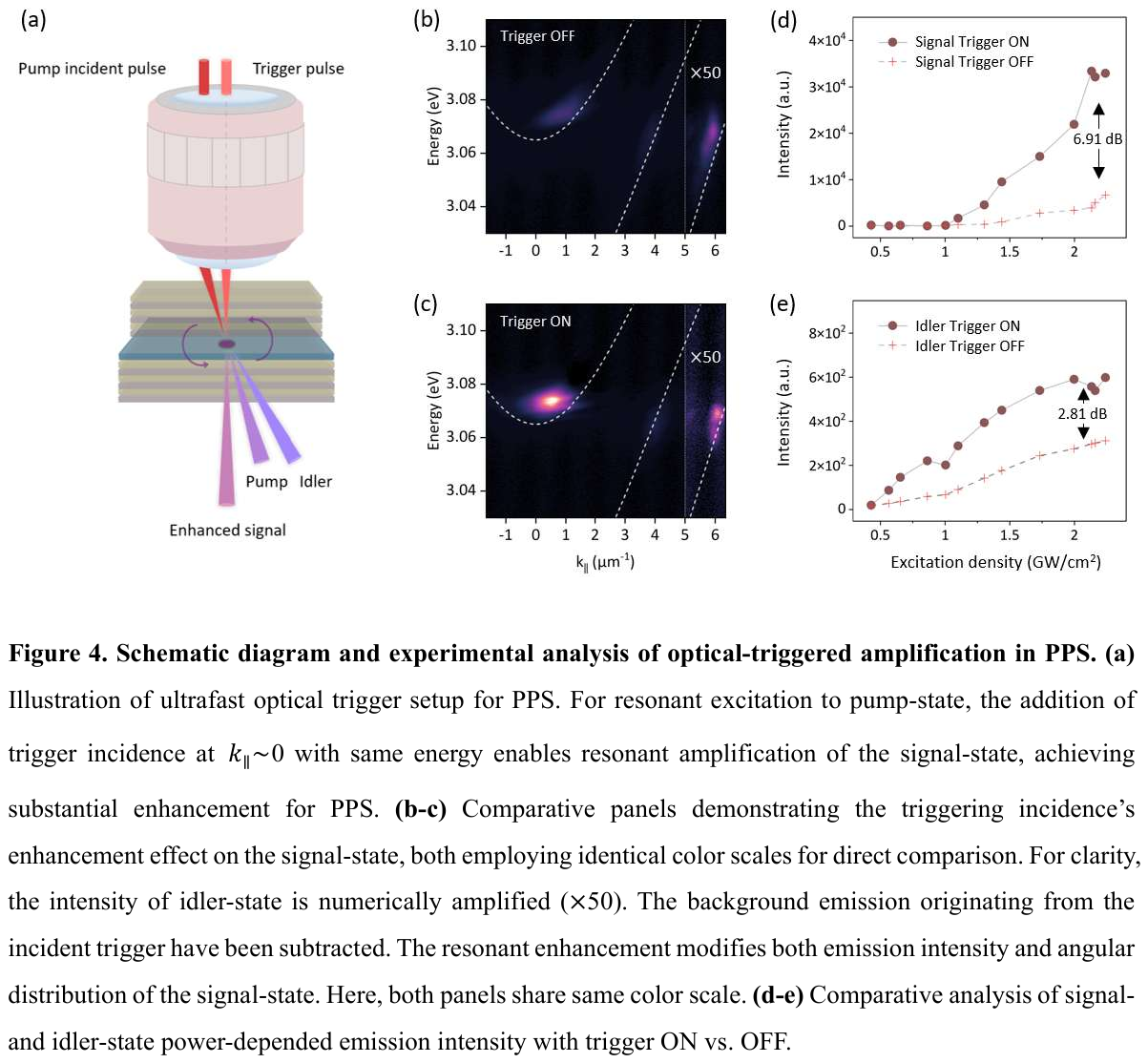}
    \label{fig_3}
\end{figure}

\clearpage{}
\newpage{}
\begin{figure}
     \centering
        \centering  \includegraphics[width=1\textwidth]{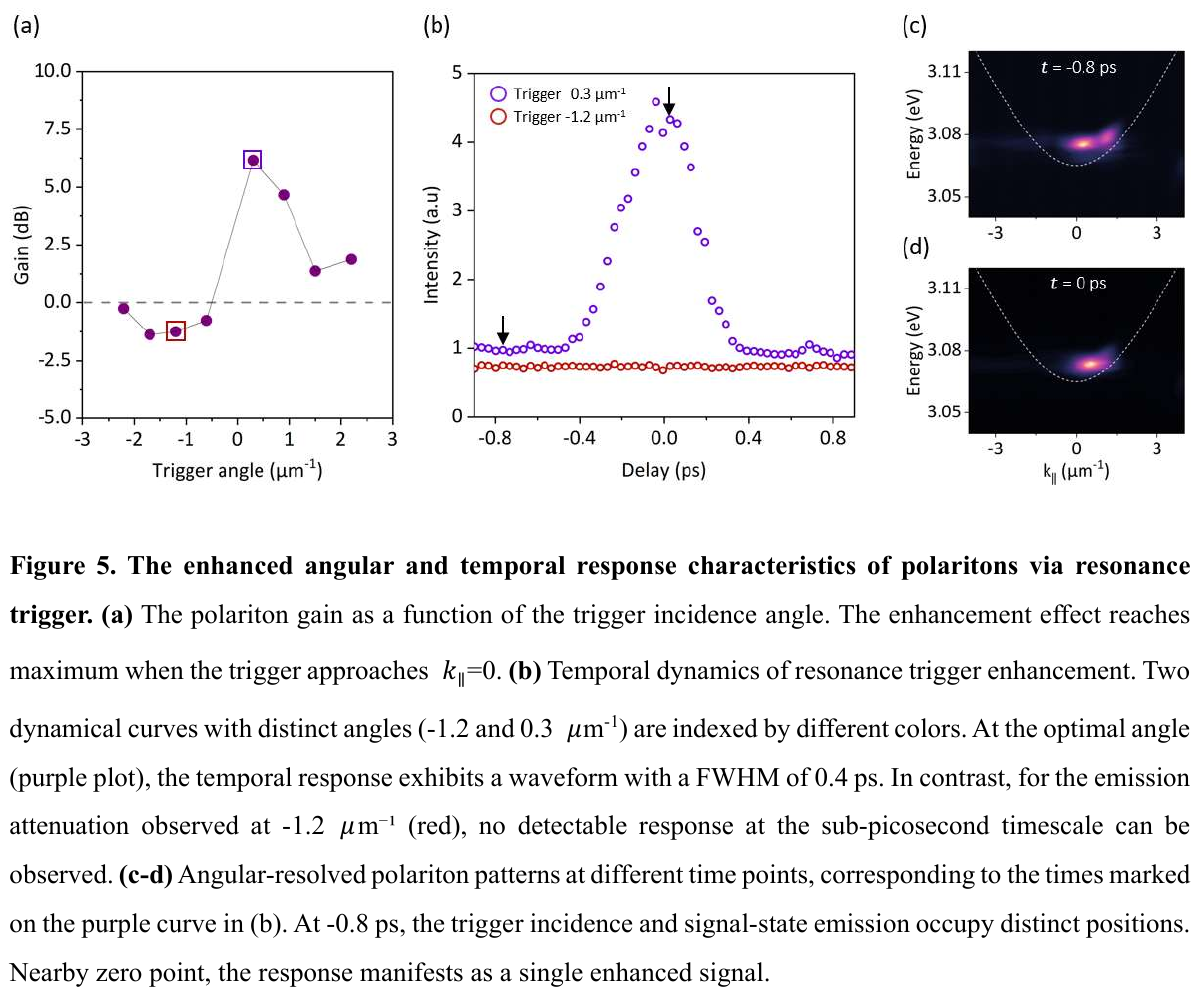}
    \label{fig_4}
\end{figure}

\newpage{}
\clearpage{}

\newpage{}
\clearpage{}

\setlength\parskip{0.25em}

\begin{spacing}{1.125}

\section*{Supplementary Information}

\section{Experimental Setup}

Femtosecond laser pulses are split by a beam splitter (BS) into two beams. Stronger beam: Serves as the resonant excitation beam for polaritons, driving the polariton parametric scattering (PPS) process via two-photon absorption (TPA). Weaker beam: Acts as the trigger signal for optical modulation experiments of PPS. The relative time delay between the two pulses is controlled by motorized translation stage. The aperture D1 confines the incident beam in k-space. Subsequently, the laser pulses are focused by the objective lens O1 and vertically incident on the ZnO microcavity. The emission signals from the microcavity are collected by another objective lens O2 and directed through a 4f imaging system for k-space resolution.The signals are separated by a second BS: One path is routed to a spectrometer with silicon charge-coupled device to acquire angle-resolved spectra. The other path is captured by a beam profiler for analyzing 2D k-space signals.
Additionally, a He-Cd laser is introduced for non-resonant excitation of the ZnO microcavity, enabling characterization of its intrinsic optical properties

 \begin{figure}[ht!]
		\centering
		\includegraphics[width=1\linewidth]{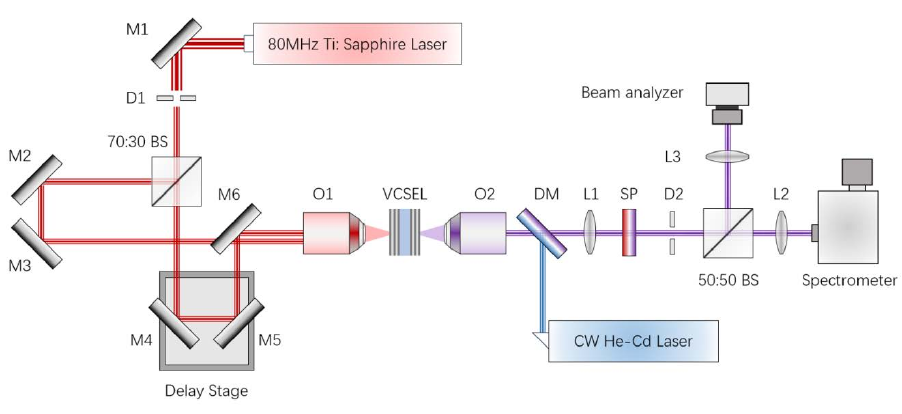}
		\caption{Schematic drawing of experimental setup. BS: beam splitter. D1 and D2: diaphragm. M1-M6: mirrors. O1 and O2: objectives. L1 and L2: lenses. P1 and P2 are polarizers. DM: dichroic mirror. SP: short-pass filter.} 
		\label{Fig_S1}
	\end{figure}

\vspace{5mm}
\section{Characterization in k-space of incident laser}

The angle-resolved spectrum of the k-space modulated incident laser beam is shown in Figure S2. The beam occupies the region of $k_{ \| } =2.5 \sim 4\mu m ^ { - 1}$ in k-space, corresponding to the pump state position in the PPS process.

 \begin{figure}[ht!]
		\centering
		\includegraphics[width=0.5\linewidth]{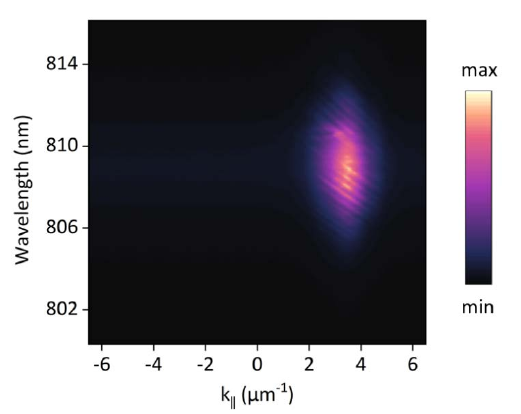}
		\caption{ Angle-resolved spectrum of incident laser} 
		\label{Fig_S2}
	\end{figure}

\vspace{5mm}
\section{Description of exciton-polaritons in planar ZnO-microcavity}

In a planar microcavity with a Fabry-Pérot type structure composed of distributed Bragg reflectors (DBRs), the dispersion relation of cavity photons is given by
\[
E_c\left(k\right)=\frac{\hbar c}{n_c}\sqrt{k_z^2+k_\parallel^2}
\]
where, for a fixed longitudinal mode order $m$, $k_z=m\pi/L_c$ is constant. Consequently, $E_c\left(k_\parallel\right)$ exhibits an approximately parabolic dependence on the in-plane wave vector $k_\parallel$ (or equivalently, the incidence angle).

Considering the exciton--photon coupling in the microcavity, the system Hamiltonian in the Hopfield model can be written as~
\[
H=E_c\left(k\right)a_k^\dag a_k+E_e\left(k\right)b_k^\dag b_k+\frac{\Omega}{2}\left(a_k^\dag b_k+b_k^\dag a_k\right)
\]
where $a_k^\dag$ is the creation operator for a cavity photon, $b_k^\dag$ is the creation operator for an exciton, and Rabi splitting $\Omega$ denotes the exciton--photon coupling strength. Upon diagonalization, the upper and lower polariton modes are expressed as
\[
p_{k,U}=X_ka_k-C_kb_k,\quad p_{k,L}=C_ka_k+X_kb_k
\]
where $X_k$ and $C_k$ are the Hopfield coefficients representing the excitonic and photonic fractions of the polariton, respectively. They satisfy the normalization condition
\[
\left|X_k\right|^2+\left|C_k\right|^2=1
\]
The analytical expressions for the Hopfield coefficients are
\[
\left|C_k\right|^2=\frac{1}{2}\left[1-\frac{\Delta E_k}{\sqrt{\Delta E_k^2+\Omega^2}}\right],\quad \left|X_k\right|^2=\frac{1}{2}\left[1+\frac{\Delta E_k}{\sqrt{\Delta E_k^2+\Omega^2}}\right]
\]
with the detuning defined as $\Delta E_k=E_c(k)-E_e(k)$. The eigenenergies of the upper polariton (UP) and lower polariton (LP) branches are then
\[
E_{\mathrm{UP},\mathrm{LP}}\left(k\right)=\frac{1}{2}\left[E_c+E_e\pm\sqrt{\left(E_c-E_e\right)^2+\Omega^2}\right]
\]
Considering a real system, with the cavity mode loss rate $\gamma_C$ and the exciton nonradiative relaxation rate $\gamma_X$, the dispersion of exciton-polariton can be written as
\[
E_{\mathrm{UP},\mathrm{LP}}\left(k\right)=\frac{1}{2}\left[E_c+E_e+i\left(\gamma_C+\gamma_X\right)\pm\sqrt{\left(E_c-E_e+i\left(\gamma_C-\gamma_X\right)\right)^2+\Omega^2}\right]
\]
It should be noted that at room temperature, the interaction between polaritons and phonons via the excitonic component is particularly strong, resulting in a much shorter lifetime for the UP branch. Consequently, polaritons rapidly relax from the exciton reservoir to the LP branch through many-body scattering processes. Under the room-temperature conditions of this work, the UP branch is almost invisible in the spectral measurements, and the observed signal is dominated by the LP branch.

For planar ZnO microcavity in this study, the exciton energy of ZnO at 300 K is 3.31~eV. The fitted Rabi splitting energies for LP1, LP2, and LP3 are 74~meV, 129~meV, and 47~meV, respectively.

\vspace{5mm}
\section{Description of PPS Simulation by Gross–Pitaevskii equation (G-P eq.)}

In order to elucidate the PPS process, the system is modeled using coupled G-P eq. for the signal, pump, and idler state polariton fields $\psi_S$, $\psi_P$, and $\psi_I$, respectively:
\begin{align*}
i\hbar\frac{\partial\psi_S}{\partial t} & =\left[{\hat{E}}_S\left(k_S\right)+g_S\left|\psi_S\right|^2-\frac{i\hbar}{2}\gamma_{\mathrm{PPS}}\right]\psi_S+\alpha_c\psi_P^2\psi_I^\ast \\
i\hbar\frac{\partial\psi_P}{\partial t} & =\left[{\hat{E}}_P\left(k_P\right)+g_P\left|\psi_P\right|^2-\frac{i\hbar}{2}\gamma_P\right]\psi_P+2\alpha_c\psi_P^\ast\psi_S\psi_I+F_P \\
i\hbar\frac{\partial\psi_I}{\partial t} & =\left[{\hat{E}}_I\left(k_S\right)+g_I\left|\psi_I\right|^2-\frac{i\hbar}{2}\gamma_{\mathrm{PPS}}\right]\psi_S+\alpha_c\psi_P^2\psi_S^\ast
\end{align*}
The PPS process satisfies both energy and momentum conservation:
\[
{\hat{E}}_S\left(k_S\right)+{\hat{E}}_I\left(k_I\right)=2{\hat{E}}_P\left(k_P\right),\quad k_S+k_S=2k_P
\]
The pump term $F_P$ is modeled in $E$-$k$ space as a Gaussian elliptical profile centered at $\left(k_P,E_P\right)$:
\[
F_P=F_0\exp{\left[-\frac{\left(k-k_p\right)^2}{A_k^2}-\frac{\left(E-E_p\right)^2}{A_E^2}\right]}
\]
The kinetic energy operators ${\hat{E}}_{S,P,I}\left(k\right)$ are derived from the dispersions of the corresponding lower polariton (LP) branches:
\[
{\hat{E}}_{S,P,I}\left(k\right)=\frac{1}{2}\left\{{\hat{E}}_{C1,C2,C3}\left(k\right)+E_X+i\left(\gamma_C+\gamma_X\right)-\sqrt{\left[{\hat{E}}_{C1,C2,C3}\left(k\right)-E_X+i\left(\gamma_C-\gamma_X\right)\right]^2+{\Omega_{1,2,3}}^2}\right\}
\]
where ${\hat{E}}_{C1,C2,C3}\left(k\right)$ are the bare cavity photon dispersions for the three LP branches, $E_X$ is the exciton energy, $\gamma_C$ and $\gamma_X$ are the photon and exciton decay rates, and $\Omega_{1,2,3}$ are the Rabi splitting energies.

In the simulations, the dispersions of the signal and pump states are set directly, while the idler dispersion is obtained from the conservation relations. By varying the $E$-$k$ positions of the signal and pump, the optimal energy-momentum matching conditions are identified. All three fields are initially prepared as Gaussian wave packets. The scattering decay rates $\gamma_{\mathrm{PPS}}$ of the signal and idler states are set to be equal, and the differences among ${\hat{E}}_S\left(k_S\right)$, ${\hat{E}}_P\left(k_P\right)$ and ${\hat{E}}_I\left(k_I\right)$ are kept small. Furthermore, the nonlinear interaction terms $g\left|\psi\right|^2$ have only a minor influence on the PPS threshold, resulting in the signal and idler reaching threshold at nearly the same pump power—consistent with experimental observations. Our theoretical simulations are based on the numerical solution of the coupled generalized G-P eq., with the following parameters: $E_X = 3.31\ \mathrm{eV}$, $m_c = 3.2\times10^{-5}\ m_e$, $\gamma_C = 1\times10^{-1}\ \mathrm{ps}^{-1}$, $\gamma_X = 1\times10^{-2}\ \mathrm{ps}^{-1}$, $g_{S,P,I} = 0.5\ \mu\mathrm{eV}\cdot\mu\mathrm{m}$, $\alpha_c = 0.25\ \mu\mathrm{eV}\cdot\mu\mathrm{m}$, $A_k = 0.75\ \mu\mathrm{m}^{-1}$, $A_E = 8\ \mathrm{meV}$.

\end{spacing}

\end{document}